\documentclass[prb,twocolumn,superscriptaddress,nofootinbib,noshowpacs,preprintnumbers,longbibliography,floatfix]{revtex4-1}
\usepackage[english]{babel}
\usepackage[utf8]{inputenc}
\usepackage{amsfonts}
\usepackage[colorinlistoftodos, color=green!40, prependcaption]{todonotes}
\usepackage[pdftex, pdftitle={Article}, pdfauthor={Author}]{hyperref} 
\usepackage{amsthm}
\usepackage{mathtools}
\usepackage{physics}
\usepackage{xcolor}
\usepackage{graphicx}
\usepackage[left=23mm,right=13mm,top=35mm,columnsep=15pt]{geometry} 
\usepackage{adjustbox}
\usepackage{placeins}
\usepackage[T1]{fontenc}
\usepackage{lipsum}
\usepackage{csquotes}

\usepackage{comment}

\begin{document}
\title{Hinge non-Hermitian skin effect in the single-particle properties of a strongly correlated f-electron system }

\author{Robert Peters}
    \email[Correspondence email address: ]{peters.robert.7n@kyoto-u.ac.jp}
    \affiliation{Department of Physics, Kyoto University, Kyoto 606-8502, Kyoto, Japan}

\author{Tsuneya Yoshida}
    \affiliation{Department of Physics, Kyoto University, Kyoto 606-8502, Kyoto, Japan}
    \affiliation{Institute for Theoretical Physics, ETH Zurich, 8093 Zurich, Switzerland}
\date{\today} 

\begin{abstract}
Non-Hermitian systems exhibit novel phenomena without Hermitian counterparts, such as exceptional points and the non-Hermitian skin effect. These non-Hermitian topological phenomena are observable in single-particle excitations of correlated systems in equilibrium, which are described by Green's functions. In this paper, we demonstrate the appearance of the hinge non-Hermitian skin effect in the effective Hamiltonian that describes the single-particle properties of an $f$-electron system. Skin effects result in a strong sensitivity to boundary conditions, and a large number of eigenstates localize at one boundary when open boundary conditions are applied. Our system exhibits such sensitivity and hosts skin modes localized around hinges.
This hinge skin effect is induced by a non-Hermitian topology of the surface Brillouin zone. The hinge skin modes are observed for 
one-dimensional subsystems located between one pair of exceptional points in the surface Brillouin zone. This paper highlights that correlated materials are an exciting platform for analyzing non-Hermitian phenomena. 
\end{abstract}


\maketitle

\section{Introduction}
Quantum systems are often inherently open and subject to dissipative processes. Non-Hermitian operators naturally describe such open quantum systems, where energy, i.e., the eigenvalues of the Hamiltonian, are not necessarily real-valued. These complex-valued band structures offer novel topological features that do not have Hermitian equivalents.\cite{PhysRevX.8.031079,El-Ganainy2018,PhysRevX.9.041015,Ashida_2020,RevModPhys.93.015005,RevModPhys.93.015005,PhysRevLett.123.066404,PhysRevB.84.205128,PhysRevB.84.153101,PhysRevB.97.121401} These features include the exceptional point (EP)\cite{Heiss2012,Mueller_2008,PhysRevLett.120.146402,science.aar7709,Wang2022,PhysRevB.99.041406,PhysRevB.99.041202,PhysRevB.99.121101,Denner2021,PhysRevB.101.205417,PhysRevLett.126.086401} and the non-Hermitian skin effect (NHSE).\cite{PhysRevLett.77.570,PhysRevLett.121.086803,PhysRevB.99.121411,PhysRevB.99.201103,PhysRevLett.123.073601,PhysRevLett.124.056802,PhysRevB.102.205118,PhysRevB.101.195147,PhysRevB.102.201103,PhysRevLett.125.126402,Zhang2022,annurev-conmatphys-040521-033133,PhysRevLett.131.256602,Zhang2022_2,Sone_2020,Lin2023,PhysRevLett.121.026808,PhysRevLett.124.086801,PhysRevLett.116.133903,PhysRevB.97.045106} The EP represents a band touching with a degeneracy of the eigenstates, which makes the Hamiltonian non-diagonalizable. 
On the other hand, the NHSE's extreme sensitivity of eigenvalues and eigenstates originate from non-trivial point-gap topology in the bulk. The localized boundary modes under open boundary conditions are known as skin modes.
Non-Hermiticity is not limited to open quantum systems.\cite{PhysRevResearch.4.023121,PhysRevA.100.062131,PhysRevLett.123.170401,PhysRevLett.127.070402,PhysRevB.92.235310} It also appears in various other systems, such as photonic crystals,\cite{RevModPhys.91.015006,Ruter2010,Regensburger2012,Zhen2015,science.aap9859,science.aar7709,PhysRevLett.117.107402,Parto2021,Ozdemir2019,PhysRevB.104.125416,PhysRevResearch.2.013280,PhysRevB.108.035406} electric circuits,\cite{PhysRevResearch.5.043034,Liu2021,Helbig2020,PhysRevResearch.2.023265,PhysRevResearch.2.022062,PhysRevB.109.115407} and correlated systems.\cite{PhysRevLett.126.176601,PhysRevB.103.125145,PhysRevB.98.035141,PhysRevB.99.121101,Yoshida_2020,PhysRevB.101.085122,PhysRevLett.124.196401,PhysRevB.107.195149,PhysRevLett.125.227204,Rausch_2021,PhysRevB.104.L121109,PhysRevB.107.195149}

Recent studies have revealed the non-Hermitian aspects of correlated materials. 
The reason for these non-Hermitian aspects is that observable quantities are often based on single-particle or two-particle operators, which can be expressed by single-particle or two-particle Green's functions, respectively. Because single-particle states or two-particle states obtain a finite lifetime due to scattering, the effective Hamiltonians describing single-particle and two-particle Green's functions are naturally non-Hermitian in correlated materials.
Therefore, non-Hermitian effects have an impact on the single-particle and two-particle properties that are described by the corresponding Green's functions. The effective Hamiltonian describing the single-particle properties of correlated materials includes the self-energy, generally, a complex-valued non-Hermitian matrix.\cite{PhysRevB.98.035141,kozii2017nonhermitian,PhysRevB.97.041203,PhysRevLett.121.026403} The complex-valued eigenvalues of this effective Hamiltonian are related to the density of states, and the eigenstates affect single-particle observables. For example, correlation effects in heavy-fermion systems lead to the emergence of EPs and a change of the band structure near the Kondo temperature.\cite{PhysRevB.98.035141,PhysRevLett.125.227204,PhysRevB.101.085122} Furthermore, it has been demonstrated that the NHSE can occur in the single-particle Green's function.\cite{PhysRevB.107.195149,PhysRevLett.126.176601,PhysRevB.103.125145}

In Ref.~\onlinecite{PhysRevB.104.235153}, it has been demonstrated that due to the enhancement of correlation effects at the surface, localized EPs appear on the surface of a topological Kondo insulator for a specific temperature range. Because the imaginary part of the self-energy in the bulk of the material is smaller, EPs only exist on the surface but not in the bulk.
 In this paper, we utilize this observation and demonstrate that these surface EPs can lead to the emergence of a hinge NHSE in the same temperature range.
Hinge skin modes are, thereby, eigenstates that are localized along the edge of a surface, i.e., these skin modes are localized along a 1D manifold in a 3D system.\cite{Shindler2018,PhysRevLett.123.073601,Xie2021,PhysRevB.102.241202,PhysRevB.99.081302,PhysRevLett.122.076801}

The argument for the occurrence of a hinge NHSE is as follows (also see Fig.~\ref{fig:model} for a schematic depiction): (1) EPs appear in the effective single-particle Hamiltonian on the surface of the material. (2) These EPs induce a non-trivial point-gap topology for a one-dimensional subsystem of the surface Brillouin zone (BZ), i.e., paths in the surface BZ passing in between one pair of EPs.
(3) The above subsystem on the surface is topologically equivalent to the Hatano-Nelson model\cite{PhysRevLett.77.570,PhysRevB.56.8651} and thus exhibits an NHSE.
As this NHSE only occurs in a subsystem of the surface BZ, we can conclude the existence of a hinge NHSE in the effective Hamiltonian describing the single-particle properties of a correlated material. 

The remainder of this paper is organized as follows. In Sec.~\ref{sec:model}, we introduce the model and methods used. In Sec.~\ref{sec:EPs}, we prove the existence of EPs on the surface of this system. This is followed by the demonstration of the NHSE on the surface for a range of momenta in Sec.~\ref{sec:NHSE}. 
 Finally, in Sec.~\ref{sec:Conclusion}, we summarize and conclude this paper.
Appendices are devoted to the analysis of the $T=0$ case and the non-Hermitian Hamiltonian under open boundary conditions in all directions.
\section{model and method}
\label{sec:model}
\subsection{Model}
\begin{figure}[t]
\includegraphics[width=\linewidth]{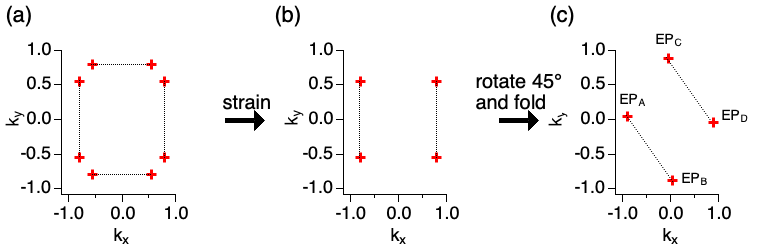}
\centering
\includegraphics[width=0.6\linewidth]{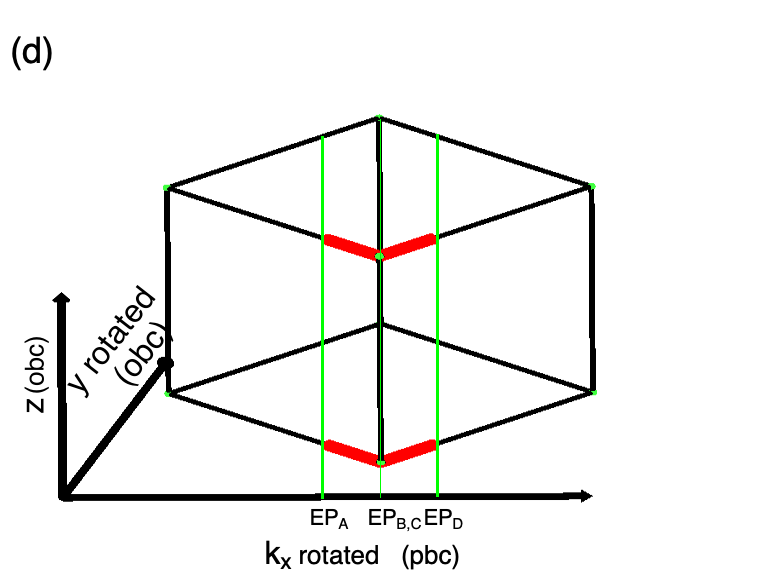}
\caption{Schematic explanation of the model and the appearance of the hinge NHSE. 
Panel~(a) shows the positions of EPs of the original cubic interacting model at finite temperatures.\cite{PhysRevB.104.235153} Two EPs with winding numbers $+1$ and $-1$ form a pair. We visualize these pairs by connecting the corresponding EPs by a black line.
Panel~(b) shows the model where the cubic symmetry is broken by strain, reducing the number of surface EPs. 
Panel~(c) shows the expected positions of EPs in the rotated, strained, and interacting model at finite temperatures.
Panel~(d) shows the hinge NHSE.
We use OBC in the $z$ and $y$ directions and PBC in the $x$ direction. Eigenstates (colored red) are expected to localize in the rotated system on the $z$ surfaces for a specific range of momenta in the surface BZ. The green lines correspond to the $k_x$ momenta at which EPs appear in panel~(c). 
\label{fig:model} }
\end{figure}

In this section, we explain the details of the used model. We start with the cubic topological Kondo insulator model, which was analyzed in Refs.~\onlinecite{PhysRevB.98.075104,PhysRevB.104.235153}. In previous works, it was shown that this model is a three-dimensional strong topological insulator with three surface Dirac cones on each surface. Furthermore, it was demonstrated that EPs in the single-particle Green's function appear on the surface for a specific range of temperatures.\cite{PhysRevB.104.235153} 
For open boundary conditions (OBC) in the $z$ direction and periodic boundary conditions (PBC) in the $x$ and $y$ directions, eight EPs surround the Dirac cone on the $z$ surface; the winding number takes $W=+1$ for four of these EPs and $W=-1$ for the others [see Fig.~\ref{fig:model}(a)].
A branch cut of eigenvalues connects an EP with winding number $W=+1$ and another EP with $W=-1$, which form a pair.
Because of the cubic symmetry of this model, most straight paths in the BZ pass in between two pairs of EPs or none. 
We modify this model here because we expect the occurrence of a hinge NHSE if a path through the BZ (a one-dimensional subsystem of the surface BZ) passes in between exactly one pair of EPs, i.e., intersects one branch cut between two EPs. In the modified model, the nearest-neighbor hopping depends on the direction, i.e., $t_x$, $t_y$, and $t_z$ are different. For the parameters chosen here,  two pairs of EPs vanish [see Fig.~\ref{fig:model}(b)].
We expect that such a modification of the model can be achieved by strain.

Next, we include a small magnetic field $B$ in the $z$ direction, which creates a gap in the Dirac cone on the $z$ surface.
We include this magnetic field to demonstrate that the appearance of the hinge NHSE is not related to the time-reversal symmetry or Hermitian topology.

The model Hamiltonian of this system reads
\begin{eqnarray}
H&=&H_0+H_{\mathrm{int}}\;,\\
H_0&=&\sum_{\vec k}\sum_{\sigma=\{\uparrow,\downarrow\}}\sum_{o=\{c,f\}}\epsilon^o_{\vec k}c^\dagger_{\vec k,\sigma,o}c_{\vec k,\sigma,o}\nonumber\\
&&+V\sum_{\vec k,\tau_1,\tau_2}c^\dagger_{\vec k,\tau_1,c}c_{\vec k,\tau_2,f}\sin k_x\sigma^x_{\tau_1\tau_2}\nonumber\\
&&+V\sum_{\vec k,\tau_1,\tau_2}c^\dagger_{\vec k,\tau_1,c}c_{\vec k,\tau_2,f}\sin k_y\sigma^y_{\tau_1\tau_2}\nonumber\\
&&+V\sum_{\vec k,\tau_1,\tau_2}c^\dagger_{\vec k,\tau_1,c}c_{\vec k,\tau_2,f}\sin k_z\sigma^z_{\tau_1\tau_2}\nonumber\\
&&+B\sum_{\vec k,\tau_1,\tau_2,o}c^\dagger_{\vec k,\tau_1,o}c_{\vec k,\tau_2,o}\sigma^z_{\tau_1\tau_2}\\
&&+\sum_{i,\sigma,o}E^o n_{i,\sigma,o}\;,
\end{eqnarray}
where the energy dispersion $\epsilon^o_{\vec k}$ is given as
\begin{eqnarray}
\epsilon^c_{\vec k}
&=&-2t_x\cos(k_x)-2t_y\cos(k_y)-2t_z\cos(k_z)\nonumber\\
&&+4t'\cdot\cos(k_x)\cos(k_y)\nonumber\\&&+4t'\cdot\cos(k_y)\cos(k_z)\nonumber\\&&+ 4t'\cdot\cos(k_x)\cos(k_z)\nonumber\\
&&+8t''\cdot\cos(k_x)\cos(k_y)\cos(k_z)\;,\\
\epsilon^f_{\vec k}&=&-0.1\epsilon^c_{\vec k}\;.\nonumber
\end{eqnarray}
The operator $c^\dagger_{\vec k,\sigma,o}$   ($c_{\vec k,\sigma,o}$) creates (annhihilates) an electron with momentum $\vec k$, spin direction $\sigma=\{ \uparrow, \downarrow \}$ in orbital $o\in\{c,f\}$.
$t_a$, $t^\prime$, and $t^{\prime\prime}$ are nearest-neighbor, next-nearest-neighbor, and next-next-nearest-neighbor hopping constants.
$V$ corresponds to a nonlocal hybridization between the $f$ and $c$ electrons, $B$ is a magnetic field in the $z$ direction, and $E^o$ are energy shifts depending on the orbital ($f$ or $c$ electrons). The local two-particle interaction for the $f$ electrons reads
\begin{eqnarray}
H_{\mathrm{int}}&=&U\sum_i n_{i,\uparrow,f}n_{i,\downarrow,f}\;.
\end{eqnarray}
We note that this Hamiltonian is Hermitian.
In this paper, we use the following parameters: $t_x/\vert t_z\vert=-1.24$, $t_y/\vert t_z\vert =-0.76$, $t_z=-1$, $t^\prime/\vert t_z\vert =t^{\prime\prime}/\vert t_z\vert =0.375$, $V/\vert t_z\vert =0.8$, $B/\vert t_z\vert =0.002$, $E^c/\vert t_z\vert =-1$, $U/\vert t_z\vert =16$, and $E^f/\vert t_z\vert =-8$. The temperature is fixed to $T/\vert t_z\vert =0.108$. We use the strength of the hopping in the $z$ direction, $\vert t_z\vert $, as the energy unit throughout this paper.
We note that the Dirac cones at $(k_x,k_y)=(0,0)$, $(\pi,0)$, and $(0,\pi)$ are gapped by the small magnetic field.
The expected positions of the EPs on the $z$ surface (OBC in the $z$ direction, PBC in the $x$ and $y$ directions) of the original model\cite{PhysRevB.104.235153}  and the modified model are shown in Fig.~\ref{fig:model}(a) and Fig.~\ref{fig:model}(b).
As explained above, we expect that the number of surface EPs will be reduced from eight to four in the interacting model at finite temperature. 
Furthermore, EPs connected by a branch cut, on which the real parts of two eigenvalues are identical, are linked by black lines in the schematic pictures of Fig.~\ref{fig:model}.
\begin{figure}[t]
    \centering
    \includegraphics[width=0.8\linewidth]{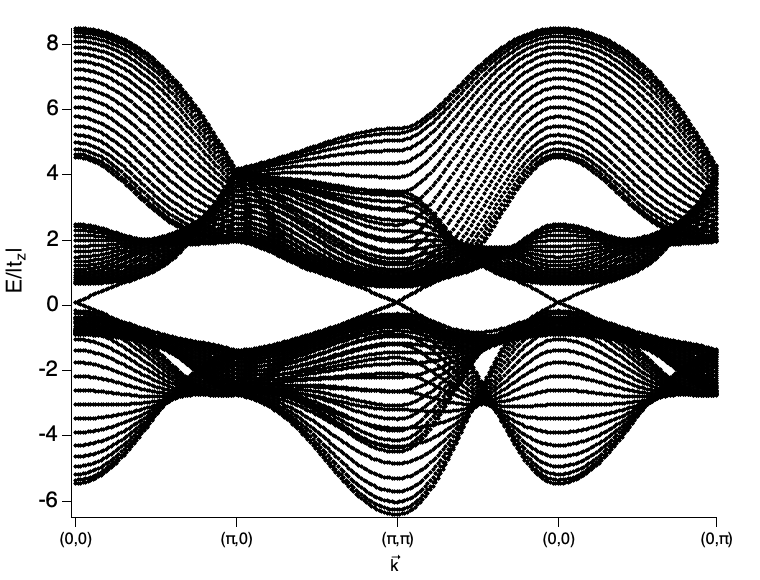}
    \caption{Noninteracting band structure of the strained rotated model. Dirac cones appear for OBC in the $z$ direction at $\vec k=(0,0)$ and $(\pi,\pi)$.}
    \label{fig:band_structure}
\end{figure}

Finally, we rotate this model in real space by $45^\circ$. This rotation leads to a change in the positions of the EPs on the surface BZ, as shown in Fig.~\ref{fig:model}(c). To obtain a periodic $45^\circ$-rotated system, we double the unit cell, resulting in a folding of the band structure.
The noninteracting band structure of this model, using the $45^\circ$-rotated geometry, is shown in Fig.~\ref{fig:band_structure}. 
After this rotation, we expect that there are straight paths along the $k_x$ or the $k_y$ direction through the BZ that pass in between exactly one pair of EPs.
These paths fulfill the necessary conditions for the appearance of the NHSE, which is the main topic of this paper.
In the results section, we will numerically demonstrate the emergence of these EPs and the appearance of the hinge NHSE.

\subsection{Method}
To solve this model, we use real-space dynamical mean-field theory (DMFT). DMFT maps each atom of a lattice on an effective quantum impurity model.\cite{PhysRevLett.62.324,Hartmann1989,RevModPhys.68.13} This mapping neglects nonlocal correlations.
Because we use OBC in the $z$ direction throughout this paper, correlation effects and, thus, self-energies depend on the distance of the atom from the surface in the $z$ direction. We thus map each layer on its own quantum impurity model.\cite{PhysRevB.93.235159} These impurity models are solved by the numerical renormalization group,\cite{RevModPhys.80.395,PhysRevB.74.245114} from which we obtain the layer-dependent self-energies at finite temperatures. Using these self-energies, we can determine new effective impurity models. This DMFT cycle is iterated until self-consistency is reached.

Having obtained self-consistent self-energies, we determine the effective Hamiltonian describing the single-particle properties by the single-particle Green's function as
\begin{eqnarray}
G(\omega)&=&(\omega+i\eta-H_{\text{eff}})^{-1},\\
H_{\text{eff}}&=&H_0+\Sigma(\omega),\label{eq:eff_Ham}
\end{eqnarray}
where $\omega$ is the frequency, $H_0$ is the noninteacting Hamiltonian, $\eta$ is an infinitesimal small, positive adiabatic factor, and $\Sigma(\omega)$ is the self-energy. We note that the self-energy depends on the layer, orbital ($f$ or $c$), and the spin direction.
Thus, $H_{\text{eff}}$ is a matrix. Furthermore, this matrix is a non-Hermitian matrix due to the imaginary part of the self-energy. 
We stress that correlations play an essential role in the non-Hermiticity of $H_{\text{eff}}$ in this heavy-fermion system.
Because the self-energy is frequency-dependent, the effective Hamiltonian also depends on the frequency. In this paper, we focus on the Fermi energy, $\omega=0$, and use $\Sigma(\omega=0)$. The single-particle properties of this system are completely determined by the eigenvalues and eigenstates of this effective Hamiltonian. The single-particle properties at the Fermi energy are thereby mostly given by the eigenvalues with $\text{Re}(E_i)\approx 0$ because then $\text{Re}(\omega-E_i)\approx 0$. Thus, we will focus on the eigenvalues of the effective Hamiltonian with a small real part.

The matrix of the effective Hamiltonian depends on the boundary conditions. We will generally use OBC with $20$ layers in the $z$ direction. To demonstrate the hinge NHSE, we will use two different kinds of boundary conditions in the $x$ and $y$ directions of the rotated system. First, we will show results for PBC in the $x$ and $y$ directions.
Second, we will show results for PBC in the $x$ direction and OBC with $40$ atoms in the $y$ direction.

\section{results}
\subsection{Energy spectrum of the rotated model}
\label{sec:EPs}
In this section, we numerically confirm that EPs emerge as illustrated in Fig.~\ref{fig:model}(c).
We analyze the energy spectrum and eigenstates of the effective Hamiltonian, Eq.~(\ref{eq:eff_Ham}), describing the single-particle Green's function in the interacting model at temperature $T/\vert t_z\vert =0.108$.
\begin{figure*}
\includegraphics[width=0.32\linewidth]{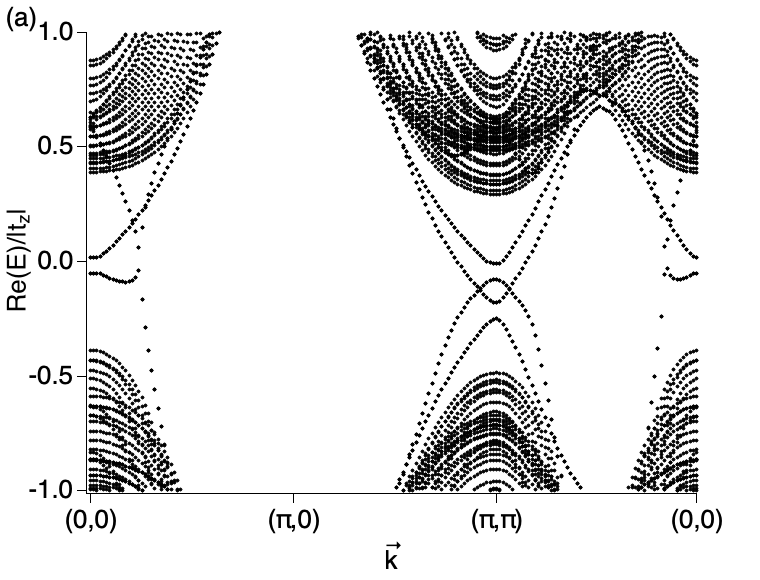}
\includegraphics[width=0.32\linewidth]{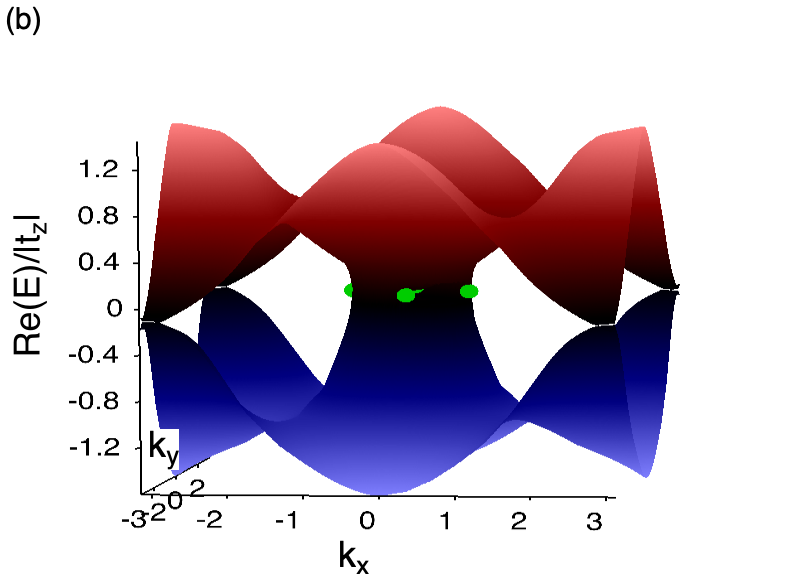}
\includegraphics[width=0.32\linewidth]{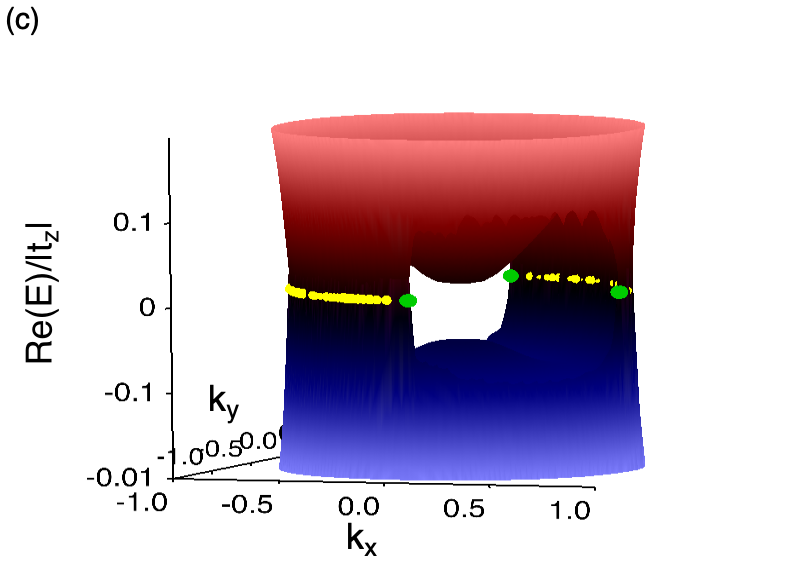}
\caption{Real part of the eigenvalues of the effective Hamiltonian at the Fermi energy. Panel (a) shows the real part of the eigenvalues close to the Fermi energy along a path in the BZ. Panel (b) shows the real part of the eigenvalues over the full BZ. Panel (c) shows a magnification of (b) around the $(k_x,k_y)=(0,0)$ close to the Fermi energy. The color in panels (b) and (c) corresponds to the real part of the energy and is just used to improve visualization. Yellow dots denote a line of momenta connected to the EPs, where the real part of the eigenvalues is degenerate. The green dots correspond to the position of the EPs, where the line of degeneracies ends. All results are for OBC in the $z$ direction and PBC in the $x$ and $y$ directions.
\label{fig:ReE}}
\end{figure*}
\begin{figure}
\includegraphics[width=\linewidth]{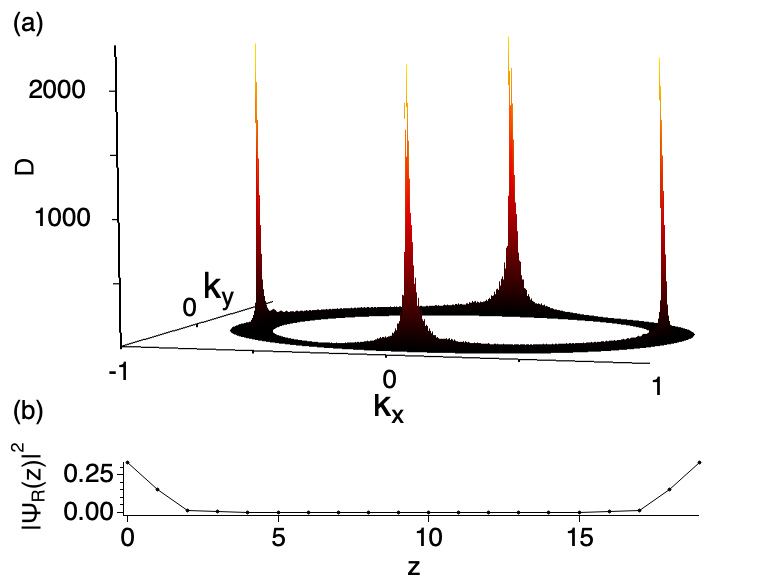}
\caption{Evidence of EPs in the surface BZ. Panel (a) shows $D$, defined in Eq.~(\ref{eq:EP}), which diverges at an EP. Panel (b) shows the square of the absolute value of the eigenstates at each EP [divergence in panel (a)] over the layer index $z$. 
The $z$ dependence, which is identical at all four EPs, demonstrates that the eigenstates are localized at the surfaces in the $z$ direction. 
\label{fig:EP}}
\end{figure}

Figure~\ref{fig:ReE} shows the real part of the eigenvalues of the effective Hamiltonian, $H_{\text{eff}}(\omega=0)$, for OBC in the $z$ direction and PBC in the $x$ and $y$ directions. From Fig.~\ref{fig:ReE}(a), it becomes clear that the self-energy arising from correlation effects has a large impact, especially around the Fermi energy. 
Low-energy eigenstates exist around $(k_x,k_y)=(0,0)$ and $(\pi,\pi)$, the original position of the Dirac cones in the rotated model, but their shape has changed due to the self-energy.
In Figs.~\ref{fig:ReE}(b) and \ref{fig:ReE}(c), we show the real part of the eigenvalues over the whole BZ and a magnification around $(k_x,k_y)=(0,0)$, respectively. We see a band surrounding $(k_x,k_y)=(0,0)$. In the magnification of this band, shown in Fig.~\ref{fig:ReE}(c), we see that this band is open and not completely surrounding the $\Gamma$ point. Additionally, we have marked the momentum values with degenerate real parts of the eigenvalue as yellow points. This line of eigenvalues with degenerate real parts suddenly ends. The end points are marked as green dots.
Such an abrupt ending of a band in the BZ, which is impossible in a noninteracting model, is the first evidence of EPs (green dots). As two EPs are connected by a band of eigenvalues with degenerate real parts in the BZ, always two EPs form a pair. In Fig.~\ref{fig:model}, we have drawn lines between EPs forming pairs.

To prove the existence of EPs in the effective Hamiltonian, we calculate a quantity describing the defectiveness of a matrix,
\begin{equation}
    D=\underset{i\neq j}{\text{max}}\frac{1}{1-\vert\langle \Psi_{R,i}\vert \Psi_{R,j}\rangle\vert}\label{eq:EP}
\end{equation}
and show it in Fig.~\ref{fig:EP}. $\vert \Psi_{R,i}\rangle$ is the $i$-th right-eigenstate of the effective Hamiltonian. 
The right-eigenstates are normalized as $\langle\Psi_{R,i}\vert\Psi_{R,i}\rangle=1$.
This property diverges if two eigenstates of the effective Hamiltonian are identical, although these eigenstates belong to different eigenvalues ($i\neq j$).
The fact that two different eigenstates are identical is evidence of an EP.
In Fig.~\ref{fig:EP}(a), we see that this property diverges at four momenta in the BZ. We can conclude that there are four EPs in the BZ. We show the positions of these EPs as green dots in Fig.~\ref{fig:ReE}. This result confirms that our expectation in Fig.~\ref{fig:model}(c) was correct.
We furthermore show the absolute value of the eigenstates at the EPs over the layer index $z$ in Fig.~\ref{fig:EP}(b). We see that these eigenstates are localized on the top and bottom surfaces in the $z$ direction.

We note that these EPs are stable while the Dirac cone is gapped due to the magnetic field and further altered by the self-energy. Small changes in the Hamiltonian, even breaking the time-reversal symmetry, do not annihilate these EPs in the effective Hamiltonian.

\subsection{Hinge skin effect induced by surface EPs}
\label{sec:NHSE}
 \begin{figure*}
\includegraphics[width=0.49\linewidth]{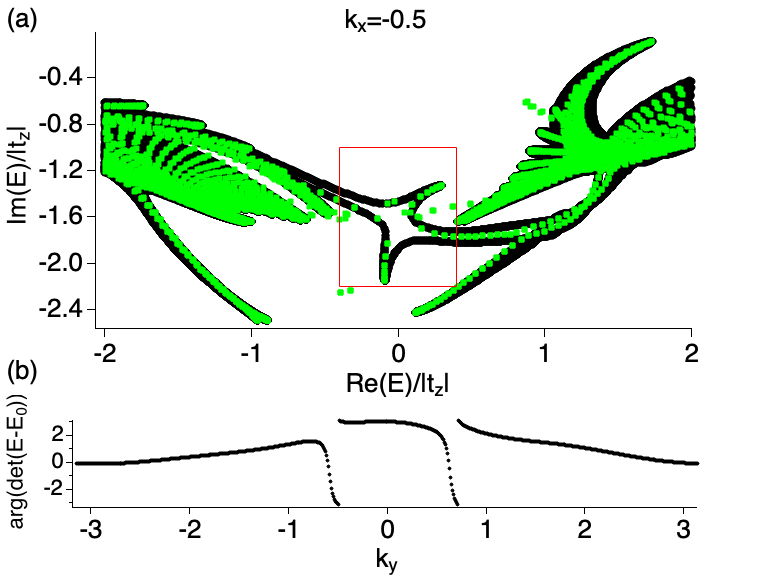}
\includegraphics[width=0.49\linewidth]{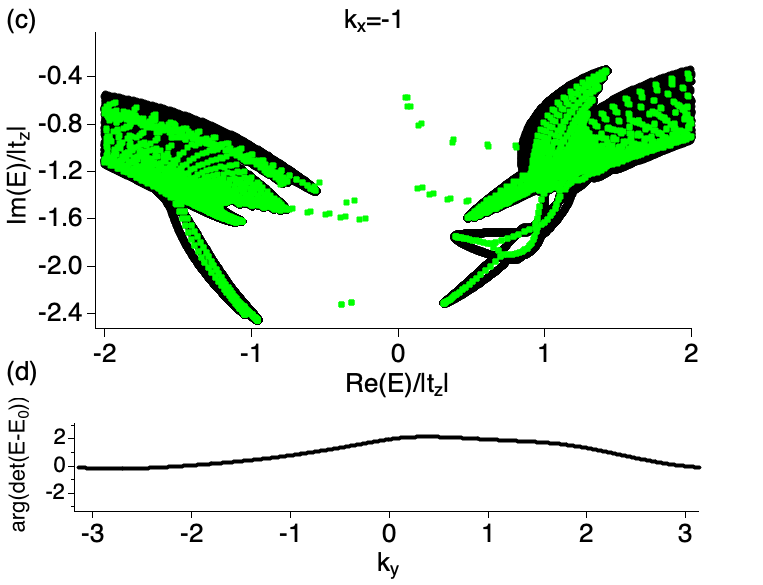}
\caption{Comparison of PBC and OBC spectra in the $y$ direction and the corresponding winding around a reference energy $E_0/\vert t_z\vert=0-1.6i$. We use PBC in the $x$ direction.
Panel (a) compares the PBC spectrum (black) with the OBC spectrum (green) for $k_x=-0.5$, resulting in a path through the BZ that intersects one pair of EPs. Clearly visible is a point gap around $\text{Re}(E)\approx 0$. We show in Fig.~\ref{fig:wavefunctions} the wavefunctions for all eigenstates inside the red rectangle. Panel (b) shows the winding for the reference energy $E_0$. Panel (c) shows the same quantities as (a) but for $k_x=-1$. In the PBC spectrum, there is no point gap. Panel (d) shows the absence of any winding for $k_x=-1$ using the same reference energy as in panel (b).\label{fig:pbc_vs_obc}}
\end{figure*}
\begin{figure}[t]
\includegraphics[width=\linewidth]{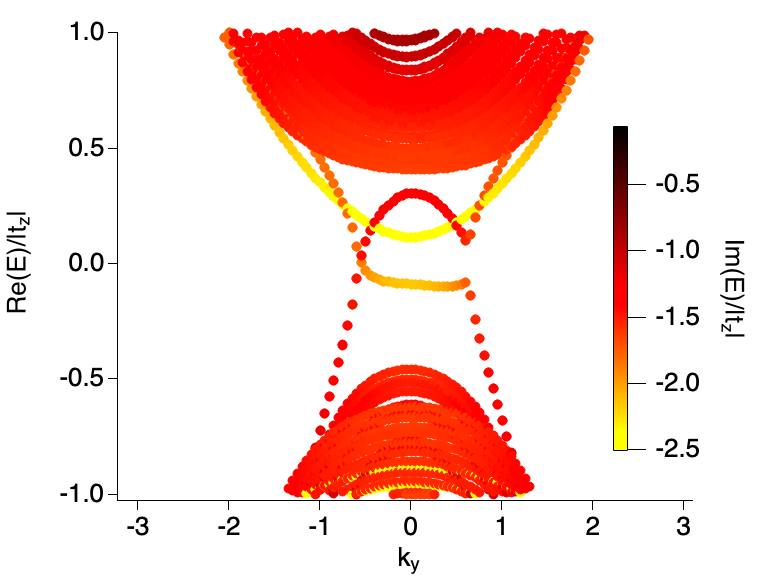}
\caption{Real parts of the eigenvalues over $k_y$ (PBC) with $k_x=-0.5$ resulting in a point gap at $\text{Re}(E)\approx 0$. The color corresponds to the imaginary part of the eigenvalues.
\label{fig:pbc_kx-05}}
\end{figure}
We demonstrate the occurrence of the NHSE for specific momenta in the system when the path through the surface BZ passes in between one pair of EPs.

\begin{figure}[t]
\includegraphics[width=\linewidth]{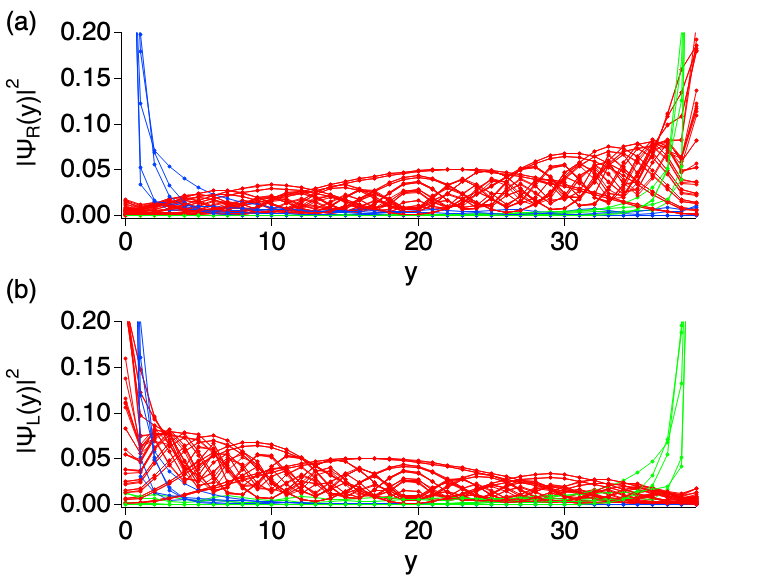}
\caption{The amplitude of the right and left eigenstates for $k_y=-0.5$ over the position in the $y$ direction. 
We show the right eigenstates, $\vert \Psi_R(y)\vert^2$, in panel (a) and the left eigenstates, $\vert \Psi_L(y)\vert^2$, in panel (b).
The red eigenstates correspond to the skin modes, where the magnitude of the right (left) eigenstates increases towards the right (left) side. Green and blue eigenstates correspond to localized states remaining from the Dirac cones of the Hermitian noninteracting model. The wave functions are normalized as $\sum_y\vert\Psi_{R/L}(y)\vert^2=1$.
\label{fig:wavefunctions}}
\end{figure}
In Fig.~\ref{fig:pbc_vs_obc}, we show the eigenvalues of the effective Hamiltonian using PBC in the $x$ direction and compare the results obtained using OBC and PBC in the $y$ direction.
In Fig.~\ref{fig:pbc_vs_obc}(a) and \ref{fig:pbc_vs_obc}(b), we fix the momentum in the $x$ direction to $k_x=-0.5$, specifying a path that intersects a branch cut between a pair of EPs when changing the $y$ direction.
In Fig.~\ref{fig:pbc_vs_obc}(c) and \ref{fig:pbc_vs_obc}(d), we use $k_x=-1$, specifying a path that does not intersect a branch cut between two EPs. In Fig.~\ref{fig:pbc_vs_obc}(a) and \ref{fig:pbc_vs_obc}(c), we show the imaginary part of the eigenvalues over the real part. The black points denote the PBC spectrum, and the green points denote the OBC spectrum with $40$ sites in the $y$ direction. In Fig.~\ref{fig:pbc_vs_obc}(b) and \ref{fig:pbc_vs_obc}(d), we furthermore show the winding of the spectrum around a reference energy $E_0( \in \mathbb{C})$ defined as
\begin{equation}
    \omega=\text{arg}\left[\text{det}(H_{\text{eff}}(k_y)-E_0\text{I})\right],
\end{equation}
which is related to the winding number as
\begin{equation}
W= \oint \frac{dk_y}{2\pi} \cdot \partial_{k_y} \text{arg} \left( \text{det}[ H_{\text{eff}}(k_y) -E_0 \text{I} ] \right).
\end{equation}

For $k_x=-0.5$ [see Fig.~\ref{fig:pbc_vs_obc}(a)], we see that there are PBC bands connecting the bulk bands for $\text{Re}(E)/\vert t_z\vert<-0.4$ with the bulk bands $\text{Re}(E)/\vert t_z\vert >0.4$. These bands clearly form a loop structure, i.e., a point gap opens. 
We have confirmed that the eigenstates forming this loop structure for $\vert \text{Re}(E)\vert/\vert t_z\vert<0.4$ are located on the $z$ surface and that the bottom and top layers contribute to these eigenstates. The layer dependence of the eigenstates in this region is thereby similar to the eigenstates shown in Fig.~\ref{fig:EP}(b).
To confirm the topological point gap, we calculate the winding in Fig.~\ref{fig:pbc_vs_obc}(b) for a reference energy ($E_0/\vert t_z\vert=-1.6i$) located inside this point gap. We see that the winding number becomes $-2$ for this reference energy.
This winding number indicates a winding of eigenvalues on the top and bottom $z$ surfaces. We note that the effective Hamiltonian fulfills reflection symmetry in the $z$-direction. A winding number of $-2$ is thus consistent with the fact that these eigenstates are located on the bottom and top surfaces and that both surfaces contribute.
Having confirmed the existence of a nontrivial point-gap topology in the energy spectrum, we expect the emergence of the NHSE.\cite{PhysRevLett.124.086801} Analyzing the OBC spectrum for $k_x=-0.5$, we see that this loop structure collapses to a single line in the energy spectrum. This sensitivity of the energy spectrum on the boundary conditions is one of the characteristics of the NHSE.

Before analyzing the OBC eigenstates for $k_x=-0.5$, we show the data for  $k_x=-1$, which corresponds to a path that does not pass in between a pair of EPs.
In Fig.~\ref{fig:pbc_vs_obc}(c), we see that for $k_x=-1$ bulk bands above $\text{Re}(E)/\vert t_z\vert>0.5$ are not connected to bulk bands $\text{Re}(E)/\vert t_z\vert<-0.5$.
Accordingly, the winding around the reference energy ($E_0/\vert t_z\vert=-1.6i$) becomes zero. Thus, we expect the absence of the NHSE. 
However, for OBC, there are several states for $-0.4<\text{Re}(E)/\vert t_z\vert<0.4$. 
These eigenstates are unrelated to the NHSE but originate in the Hermitian noninteracting Hamiltonian, as we demonstrate in Appendix~\ref{appendix_A}.

For a better understanding of the nontrivial point gap topology,
we show $\text{Re}(E)$ over $k_y$ for $k_x=-0.5$ (PBC) in Fig.~\ref{fig:pbc_kx-05} again. The color corresponds to the imaginary part of each eigenvalue. We clearly see a crossing of two bands at $\text{Re}(E)\approx 0$ and $k_y\approx -0.8$. This crossing, i.e., a degeneracy of the real part, corresponds to the above-mentioned branch cut connecting two EPs, visualized as yellow dots in  Fig.~\ref{fig:ReE}(c). Such a crossing exists only for paths that pass in between a pair of EPs.
Thus, there exists only one crossing because this crossing is generated by the EPs. 
From the color in Fig.~\ref{fig:pbc_kx-05}, we see that the imaginary parts of these two crossing bands are different. Thus, these two intersecting bands form the nontrivial point-gap topology in Fig.~\ref{fig:pbc_vs_obc}(a). 

Finally, we prove the NHSE on the $z$ surface for $k_y=-0.5$ by showing the right and left eigenstates for $-0.4<\text{Re}(E)/\vert t_z\vert<0.4$ over the position in the $y$ direction (OBC)  in Fig.~\ref{fig:wavefunctions}. 
The plotted eigenstates are restricted to those inside the red rectangle in Fig.~\ref{fig:pbc_vs_obc}(a).
Figure~\ref{fig:wavefunctions}(a) [\ref{fig:wavefunctions}(b)] shows the right [left] eigenstates.
A large number of right eigenstates (red color) have an increased absolute value, $\vert\Psi_R(y)\vert^2$, on the right side when compared to the left side. The absolute value of these right eigenvectors has a positive slope over the $y$ position.
The left eigenvectors for the same eigenvalues [colored red in Fig.~\ref{fig:wavefunctions}(b)] have a negative slope. We can thus conclude that a large number of right eigenstates are mainly localized on the right side, and the left eigenstates are mainly localized on the left side of the system with open boundaries in the $y$ direction.
This is the expected behavior of the NHSE, in which a large number of left and right eigenstates localize on the opposite edges of the system.
We stress that these skin modes localize around a hinge of the three-dimensional system; the localization occurs on the $z$ surface for a specific range of $k_x$ values.

Besides these skin modes, there are other eigenstates that localize at the boundaries in the $y$ direction.
These eigenstates maintain their position of localization when changing between right and left eigenstates. 
In Fig.~\ref{fig:wavefunctions}, both right and left eigenstates denoted by green (blue) lines are localized around $y=40$ ($y=0$).
These eigenstates arise from the original Hermitian Hamiltonian and are unrelated to non-Hermiticity, as shown in Appendix~\ref{appendix_A}.
These states can be seen in Fig.~\ref{fig:pbc_vs_obc} as green dots appearing seemingly unrelated to the PBC band structure (black).

\section{Conclusions}
\label{sec:Conclusion}
We have demonstrated the appearance of the hinge NHSE in the effective Hamiltonian, which describes the single-particle Green's function. The used model Hamiltonian is derived from a three-dimensional strong topological Kondo insulator under magnetic field and strains.
By showing the energy spectrum along a path passing in between one pair of EPs, we have demonstrated the existence of a nontrivial point-gap topology necessary for the emergence of the NHSE.
We furthermore have shown that a large number of right eigenstates localize at the right boundary for OBC in the $y$ direction and PBC in the $x$ direction, while the corresponding left eigenstates localize at the left boundary. All these skin modes are localized on the boundaries of the top and bottom $z$ surfaces. 
Thus, we have demonstrated the emergence of the hinge NHSE in the effective Hamiltonian, describing the single-particle properties of this strongly correlated system.

Furthermore, in  Appendix~\ref{appendix_B}, we observe the existence of localized eigenstates of the effective Hamiltonian under OBC in all directions. These eigenstates also localize along the hinges of the bottom and top surfaces or at the corners of the system. However, the topological origin of these localized modes is left as a future work.

\begin{acknowledgements}
R.P. is supported by JSPS KAKENHI No.~JP23K03300. 
T.Y. is supported by JSPS KAKENHI Grant Nos.~JP21K13850 and JP23KK0247, JSPS Bilateral Program No.~JPJSBP120249925.
T.Y is grateful for the support from the ETH Pauli Center for Theoretical Studies and the Grant from the Yamada Science Foundation.
Parts of the numerical simulations in this work have been done using the facilities of the Supercomputer Center at the Institute for Solid State Physics, the University of Tokyo.
\end{acknowledgements}

\appendix
\section{Hermitian boundary states}
\label{appendix_A}
\begin{figure}[t]
\includegraphics[width=\linewidth]{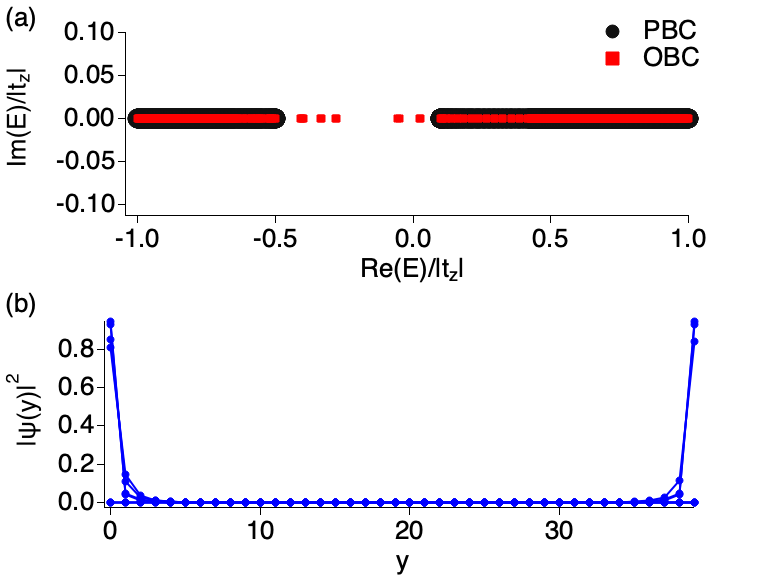}
\caption{The energy spectrum for OBC and PBC at $T=0$, for the Hermitian effective Hamiltonian.
Panel (a) shows the energy spectrum for OBC and PBC. Clearly visible is a gap in the PBC spectrum and in-gap states in the OBC spectrum.
Panel (b) shows the wave functions of these in-gap states for OBC in the $y$ direction. The wave function is normalized as
$\sum_y\vert\Psi(y)\vert^2=1$.
\label{fig:hermitian}}
\end{figure}
In this appendix, we analyze the energy spectrum and wave functions in the Hermitian case at $T=0$. At $T=0$, the imaginary part of the self-energy at the Fermi energy vanishes. Thus, the matrix describing the single-particle properties at the Fermi energy becomes Hermitian.

In Fig.~\ref{fig:hermitian}(a), we show the energy spectrum of the rotated model for OBC in the $z$ direction and PBC in the $x$ direction. We compare OBC and PBC in the $y$ direction. As expected, the imaginary part of the eigenvalues vanishes for the Hermitian matrix. Furthermore, we see that the PBC spectrum includes a gap at the Fermi energy, $Re(E)=0$. When changing to OBC in the $y$ direction, we see eigenvalues inside this gap. The wave functions of these eigenstates are shown in Fig.~\ref{fig:hermitian}(b), demonstrating that these eigenstates are localized at the boundaries in the $y$ direction. We believe these eigenstates have the same origin as the localized eigenstates in Figs.~\ref{fig:wavefunctions} that do not originate from the NHSE.

\section{Open boundary conditions}
\label{appendix_B}
\begin{figure}[t]
\includegraphics[width=\linewidth]{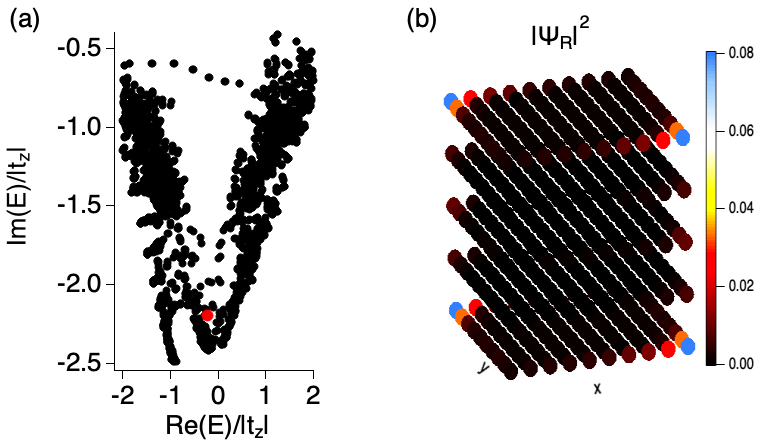}
\caption{Energy spectrum and one right eigenstate in the system with OBC in all directions. We diagonalize the effective Hamiltonian in the rotated geometry corresponding to Fig.~\ref{fig:model}(c).
Panel (a) shows the imaginary part over the real part of the energy of the effective Hamiltonian.
In panel (b), the color represents the amplitude of the right eigenstate [indicated by the red color in panel (a)] over the position in the system.
\label{fig:OBC_E}}
\end{figure}
\begin{figure}[t]
\includegraphics[width=\linewidth]{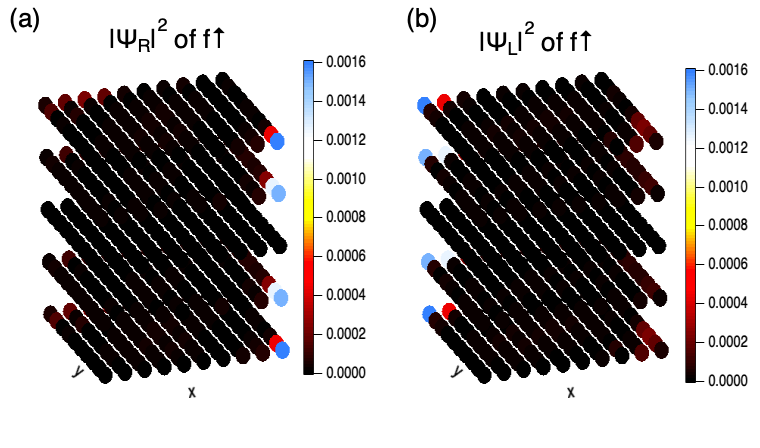}
\includegraphics[width=\linewidth]{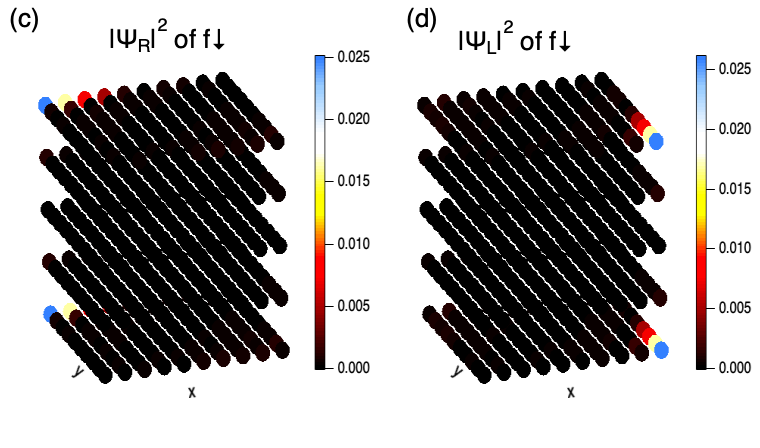}
\caption{Right ($\Psi_R$) and left eigenstates ($\Psi_L$) in the system with OBC in all directions. Panels (a) and (b) show the $f$-orbital, spin-up component of the 
right and left eigenstate [indicated by the red color in Fig.~\ref{fig:OBC_E}(a)]. The color indicates the amplitude over the position of the atom in the lattice.
 Panels (c) and (d) show the $f$-orbital, spin-down component of the same
right and left eigenstate.
\label{fig:OBC_f_wave}}
\end{figure}
We analyze the effective Hamiltonian in a system with OBC in all directions. We diagonalize the effective Hamiltonian (including the self-energy) in the rotated geometry, corresponding to Fig.~\ref{fig:model}(c).
The size of the system in all three directions is chosen as $(L_x,L_y,L_z)=(10,10,5)$. We note that this results in a Hilbert space dimension of $4000$, which can still be completely diagonalized. Here, we make the simplification that the self-energy only depends on the layer index in the $z$-direction, corresponding to a similar situation as in the previous sections. Then, the imaginary parts of the bottom and top layers are again larger than the imaginary parts of the middle layers, which would lead to the emergence of exceptional points on the bottom and top surfaces in a system with PBC.

In the main text, we showed the emergence of the non-Hermitian skin effect in the $y$-direction using PBC in the $x$-direction.
From the position of the exceptional points in Fig.~\ref{fig:model}(c), it is clear that there is also a skin effect in the $x$-direction when using PBC in the $y$-direction. Thus, the emergence of a non-Hermitian skin effect on the bottom and top layers in the $x$- and $y$-direction seems possible.
To confirm this, we show the energy spectrum (imaginary part over real part) of the fully open system in Fig.~\ref{fig:OBC_E}(a). Comparing these energies with Fig.~\ref{fig:pbc_vs_obc}, we see that there are eigenstates of the fully open system in the same energy region where we previously observed the non-Hermitian skin effect. In Fig.~\ref{fig:OBC_E}(b), we show the amplitude of a right eigenstate [indicated by the red color in Fig.~\ref{fig:OBC_E}(a)]  as a color over the position of the atom in the lattice. We see that this eigenstate is localized at the corners of the bottom and top surfaces. This is a corner state localizing simultaneously in the $x$- and $y$-directions of the surface layers. Contrary to the results in the main text, where the skin effect led to a localization at only one boundary, we now see a localization at two opposite boundaries. We note that many eigenstates with Re(E)$\approx 0$ show a similar behavior. The localization at two opposite boundaries may be understood by the fact that the original multiband Hamiltonian fulfills time-reversal symmetry. The non-Hermitian skin effect in systems with time-reversal symmetry leads to degenerate skin modes localized at opposite boundaries. Although, in the current system, the time-reversal symmetry is broken by a small magnetic field, the breaking of the symmetry is not very strong. Thus, in the fully open system, we can expect a mixing of eigenstates localized at opposite boundaries. 
To justify this hypothesis, we show the left and right eigenstates of the $f$-orbital with spin-up in Fig.~\ref{fig:OBC_f_wave}(a,b) and the left and right eigenstates of the $f$-orbital with spin-down in Fig.~\ref{fig:OBC_f_wave}(c,d). Although time-reversal symmetry is broken in this effective Hamiltonian, we can nevertheless expect that the spin-up and spin-down components behave approximately in the opposite way. 
We immediately see that, as expected for skin modes, these eigenstates localize only in one corner of the bottom and top surfaces. Furthermore, the localization of the right and left eigenstates is opposite, which confirms this fundamental property of skin modes.
Finally, we see that the localization of the spin-up component and the spin-down component are also opposite. As the full eigenstate shown in Fig.~\ref{fig:OBC_E}(b) consists of spin-up and spin-down components, this state is localized at two opposite corners of the bottom and surface layers. Based on this argument, we believe that the skin modes of this multiband system are localized in two opposite directions on the bottom and top layers. 

Thus, the obtained numerical data indicates the existence of skin modes that are localized at the hinges or corners of the bottom and top surfaces under OBC in all directions. However, we stress that the topological characterization of the skin modes performed in the main text is only valid for a system with mixed boundary conditions (OBC in the $x$- and $z$-directions and PBC in the $y$-direction). The topological origin of the skin modes in the fully open system is left as a future work.

\bibliography{library.bib}

\end{document}